\documentclass[12pt]{iopart}
\usepackage{graphicx}
\usepackage{epsfig}

\begin{document}

\title[QCD at non-zero density]{QCD at Non-Zero Density : Lattice Results}

\author{M P Lombardo}

\address{INFN LNF \\ I 00044 Frascati (Roma) Italy}
\ead{lombardo@lnf.infn.it}
\begin{abstract}
A  concise review of the progress of lattice calculations at
non-zero density since QM2006, with emphasis on the high baryon density,
low temperature domain. Possibilities for exploring  densities higher than
those studied by standard techniques are analysed.
The phase transitions of  cold, dense matter, where the sign problem
remains severe, are discussed in the context  of  QCD-like models and
approximations to QCD.

\end{abstract}

\section{Introduction}

\begin{figure}[b]
\includegraphics[width=0.4\textwidth,angle=270]{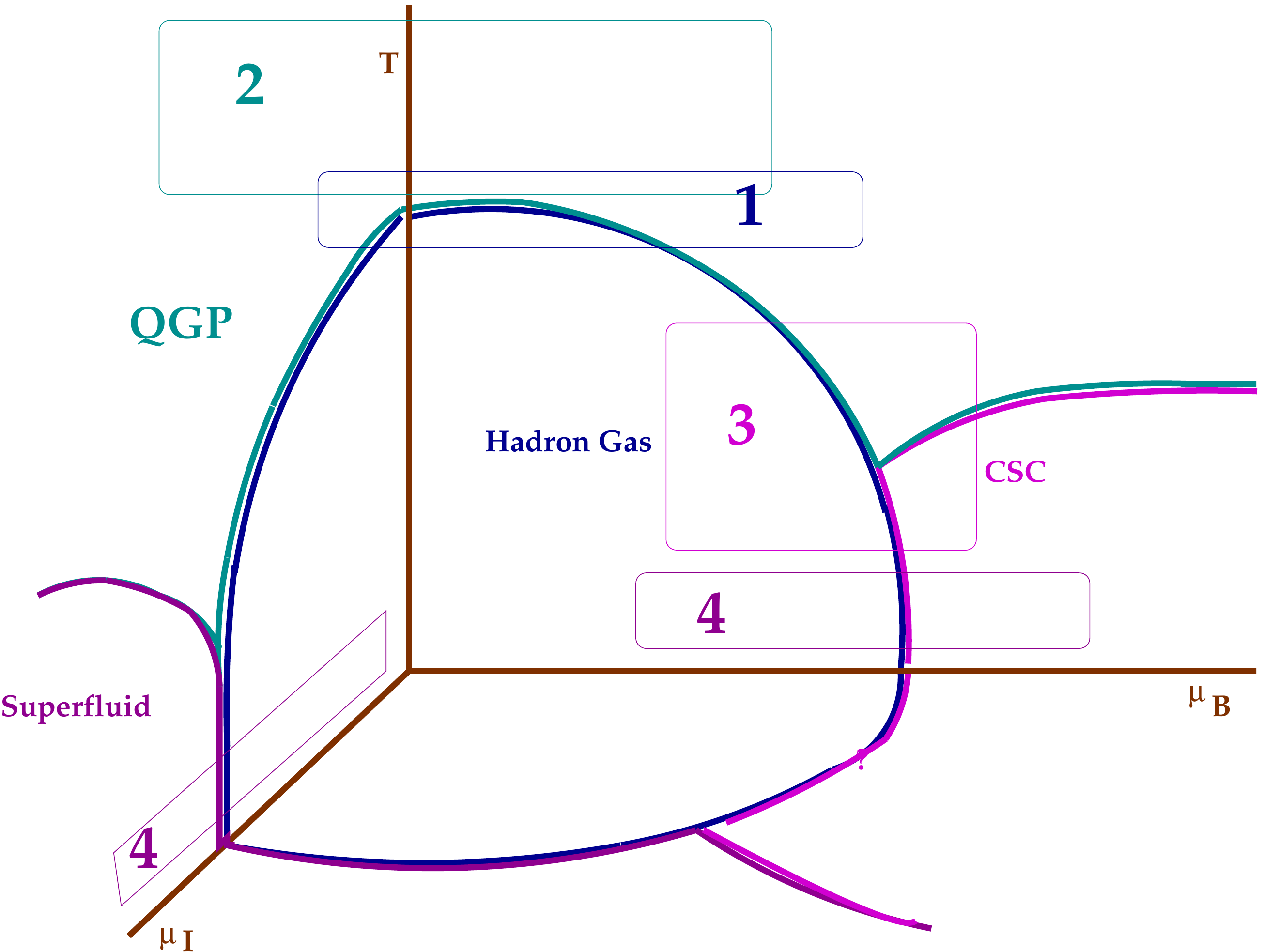}
\caption{The phase diagram of QCD in the T, $\mu_B, \mu_I$ plane.
Mature calculations are possible around $T_c$, for small $\mu_B$,
in the region  marked ({\bf 1}) in the plot, 
and in  the domain of the strongly interactive QGP ({\bf 2}).  Progress
is being made  in the parameter range of potential interest to FAIR
( {\bf 3} ),
while the physics of cold, dense matter ( {\bf 4} ) can be explored only
with a nonzero isospin density, or by use of two-color QCD.}
\label{fig:phase}
\end{figure}

The subject of this note is illustrated in Fig. 
\ref{fig:phase}, 
where I have
sketched the phase diagram of QCD in the temperature, baryochemical
potential, isospin chemical potential space. Lattice
calculations aim at a quantitative analysis of the QCD phase diagram, using 
the QCD Lagrangian as a sole input. I will 
review the progress towards this goal 
since QM2006\cite{Hatsuda:2007rt} examining the
various thermodynamic regions indicated in Fig.\ref{fig:phase}.

\section{The Critical line and the Critical Point}
A critical point in the T,$\mu_B$ plane has been predicted by model
calculations, and several lattice studies have searched for it, 
without finding a general agreement, see e.g. 
\cite{Stephanov:2007fk} for a review. 
In ref.\cite{deForcrand:2006pv}
an alternative strategy was proposed, based on the analysis of the
slope $K$ of the critical surfaces in the 
$ m_{u,d}, m_s, \mu_B$ space stemming from the 
critical  line which limits the first order transition
area of  the $m_{u,d}, m_s$ plane:
\begin{equation}
m_c(\mu) = 1 + K (\mu/T)^2
\end{equation}
A positive slope would be indicative of a critical endpoint.
$K$ was computed for three
degenerate quark masses $m_{u,d} = m_s$, giving $K= -3.3(5)$
\cite{deForcrand:2007rq}. 
\begin{figure}[t]
\vskip -9 truecm
\includegraphics[width=0.7\textwidth]{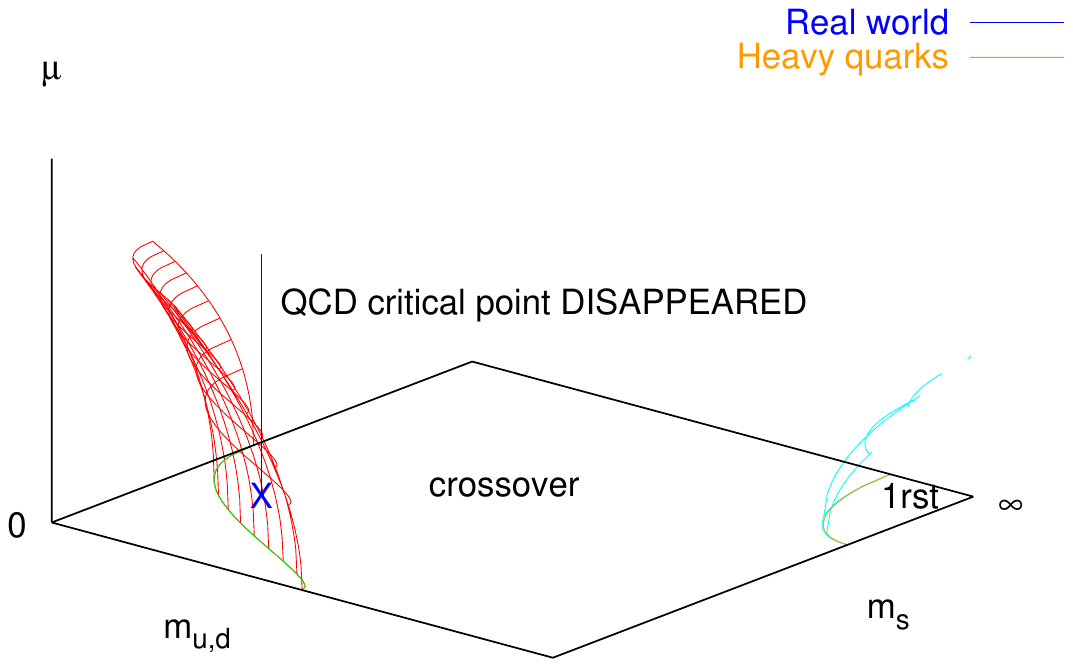}
\hskip -3 truecm
\includegraphics[width=0.7\textwidth]{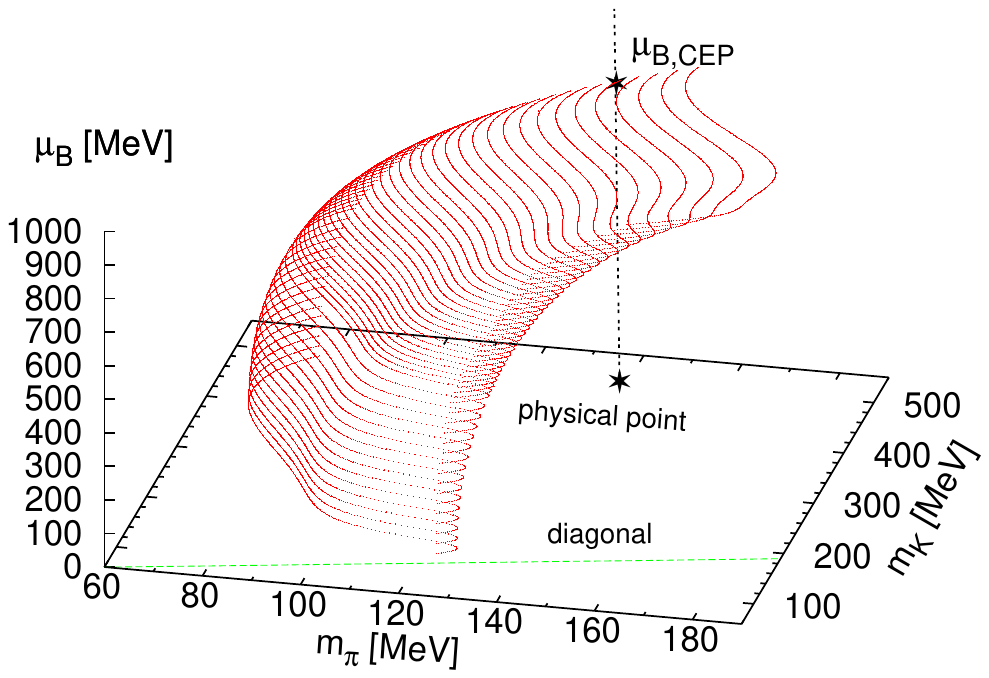}
\label{fig:njl}
\vskip -2 truecm
\hskip -2 truecm
\caption{The critical surface of QCD, if the current  results 
obtained on moderately fine lattices would
persist in the continuum limit (left, after \cite{deForcrand:2007rq});
the critical surface of a   SU(3)XSU(3) chiral model, from 
\cite{Kovacs:2006ym},  which
contains the endpoint of the chiral line in the $T, \mu_B$ plane. }
\end{figure}
An analogous calculation,.
but for isospin baryochemical potential, give $K= -3.0(1)$ 
\cite{Sinclair:2006zm}.
Close to the zero density critical temperature the difference
between effects produced by baryochemical potential
and isospin chemical potential is small 
( for instance, large $N_c$ calculation
predict differences $O(1/N_c^2)$ \cite{Toublan:2005rq}),
 so the two results
 \cite{deForcrand:2007rq, Sinclair:2006zm} 
consistently suggest
that the region of first order phase transition becomes smaller at nozero
chemical potential, at least till $\mu < 600 $ MeV. A mean
field calculation carried out in the $SU(3)_L X SU(3)_R$ NJL model
shows instead the opposite trend\cite{Kovacs:2006ym}, see Fig.
 \ref{fig:njl}, right. 
All in all we are observing qualitative 
differences between QCD and purely fermionic models at small chemical
potential. Indeed chiral symmetry is most likely broken by  
long distance forces in QCD, and by  short distance vector forces
in NJL, and these differences might well have an impact on the existence 
of the critical endpoint, see also ref. \cite{Fukushima:2008}. 

The results \cite{deForcrand:2007rq, Sinclair:2006zm}   
have been obtained on lattices with $N_t=4$.
It would be very important to confirm these predictions closer to the 
continuum limit. 
Indeed, the relevance of taking the continuum limit can be hardly
overemphasised in QCD \cite{lattice07} .

One substantial step towards the continuum limit of the critical line
has been presented at this meeting \cite{Fodor:2007pg}.
The slope of the critical line
has been computed for $N_t = 4,6,8,10$, and the claim is that there 
is a reasonable control over the $a \to 0$ extrapolation, see Fig.3.
Note that while the (pseudo)critical 
temperatures  at $\mu=0.0$  estimated from the peak of the Polyakov loop
susceptibility differs from the one estimated from the peak of the chiral
susceptibility, the slope of the two lines are the same within errorbars. 
\begin{figure}
\hskip 1.0 truecm \includegraphics[width=0.6\textwidth]{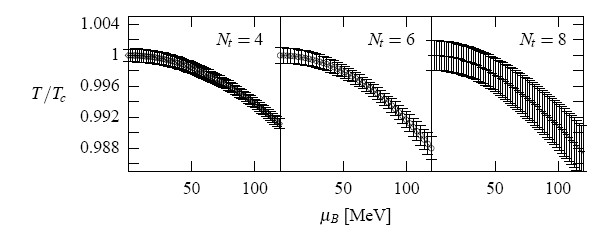}\includegraphics[width=0.3\textwidth]{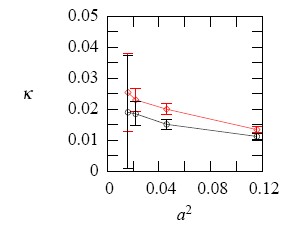}
\caption{The approach to the continuum limit of the critical line 
\cite{Fodor:2007pg}. The critical lines computed
on lattices with size $N_t=4, N_t=6, N_t=8$ are shown in the left plot, 
while the slope of the
critical line, computed from two different susceptibilities, 
as a function of the lattice spacing are summarised in the right plot.
Courtesy  C. Guse.}
\end{figure}

\section{Equation of State and Critical Behaviour}
Thermodynamics studies on the lattice are based on the 
analysis of the number
density 
$n_{u,d}(T,\mu_u,\mu_d,m_u,m_d)=
\frac{\partial  p(T,\mu_u,\mu_d)}{\partial\mu_{u,d}}$ 
and its associated susceptibilities 
\begin{equation}
\chi_{j_u,j_d}(T)=\left.
\frac{\partial^{(j_u + j_d)} p(T,\mu_u,\mu_d)}{\partial\mu_u^{j_u} \, \partial\mu_d^{j_d}}
\right|_{\mu_u=\mu_d=0}.
\end{equation}
The latter are a significant probe of the fluctuational behaviour of
the system\cite{Gavai:2005yk,Bernard:2007nm},
 and are also the Taylor coefficients of the excess pressure
$\Delta p(T,\mu_u,\mu_d) \equiv p(T,\mu_u,\mu_d) - p(T,\mu_u=0,\mu_d=0)$
which  contains information about baryon density effects in the EoS. 
By exploiting these observables 
it is was established already at the time of QM2006 
that  the hadron resonance gas model ,
where $n (T, \mu) = K(T)  \sinh (N_c \mu / T) $ 
describes well the system up to rather high temperatures
$T \simeq 0.95 T_c$ \cite{Karsch:2003zq, D'Elia:2004at}.  

\begin{figure}[b]
\begin{minipage}{\textwidth}
 \includegraphics[width=6.5cm] {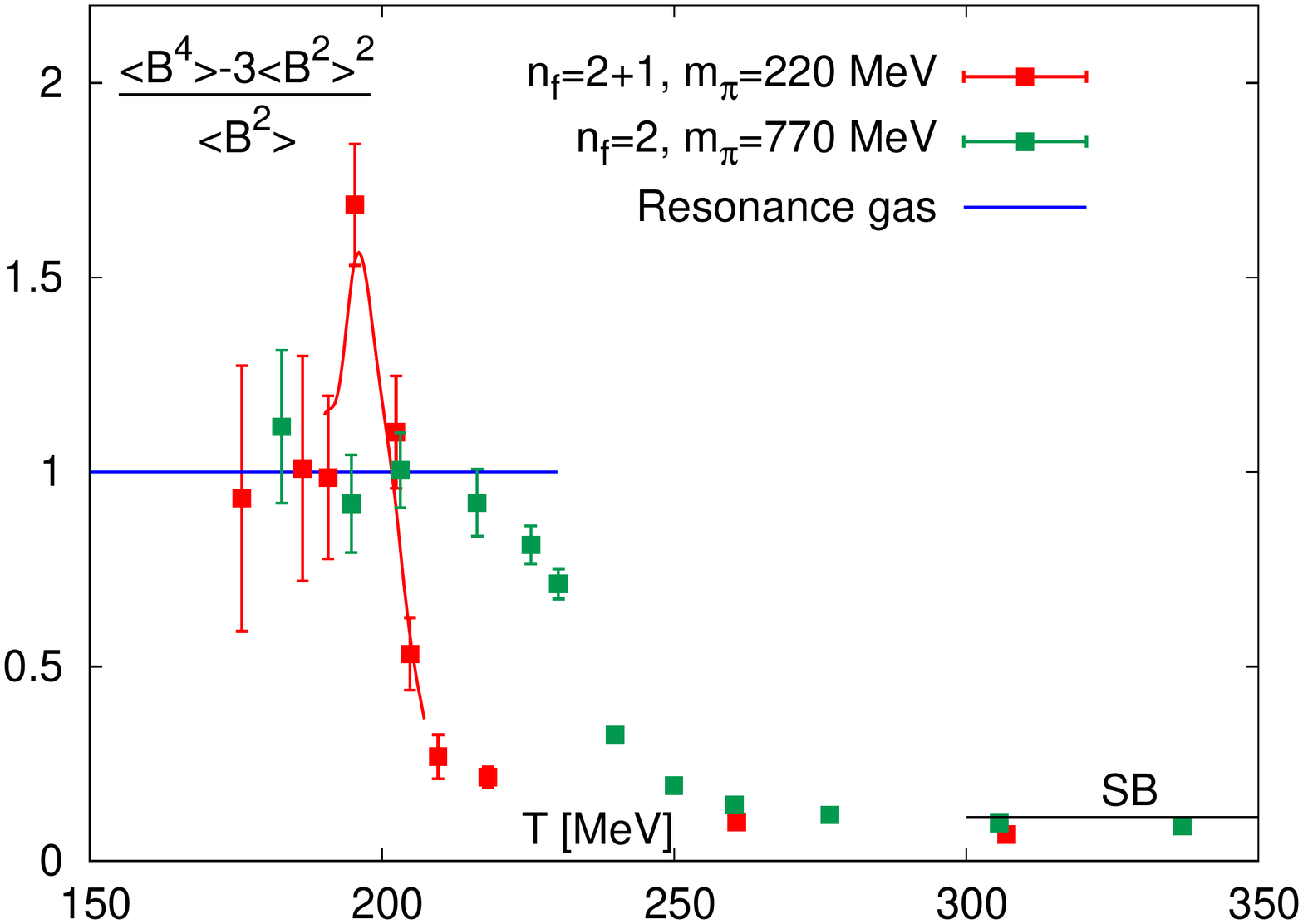}
\includegraphics[width=6.5cm] {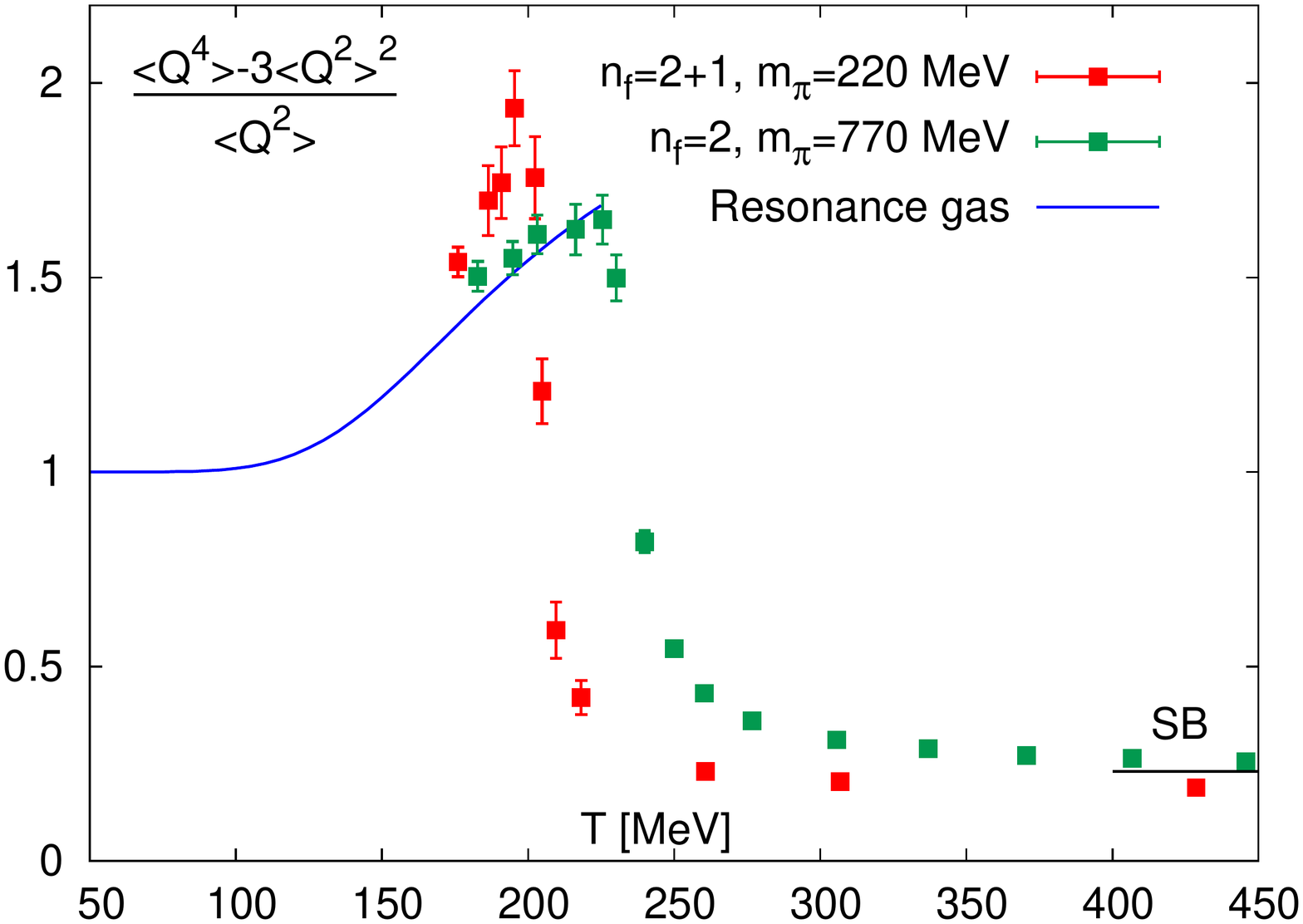}
\caption{Susceptibilities as a function of temperature \cite{Chr} 
demonstrating that the
critical region of real QCD region cannot be described by simple models.Courtesy C. Schmidt}
\label{fig:chr}
\end{minipage}
\end{figure}

For zero chemical potential, 
new results for the susceptibilites have been presented by the RBC-Bielefeld
collaboration\cite{Chr}.
They confirm the hadron gas behaviour at low temperature,
and the approach to the free gas at high temperature. Most important,
they note that these
behaviours are not smoothly connected, rather
there is a peak in between associated with the critical behaviour,
see Fig. \ref{fig:chr}. 

\begin{figure}[t]
\vskip -11 truecm
\begin{minipage}{24 truecm}
\begin{minipage}{11 truecm}
\hskip -1.0 truecm
\includegraphics[width=1.2\textwidth]{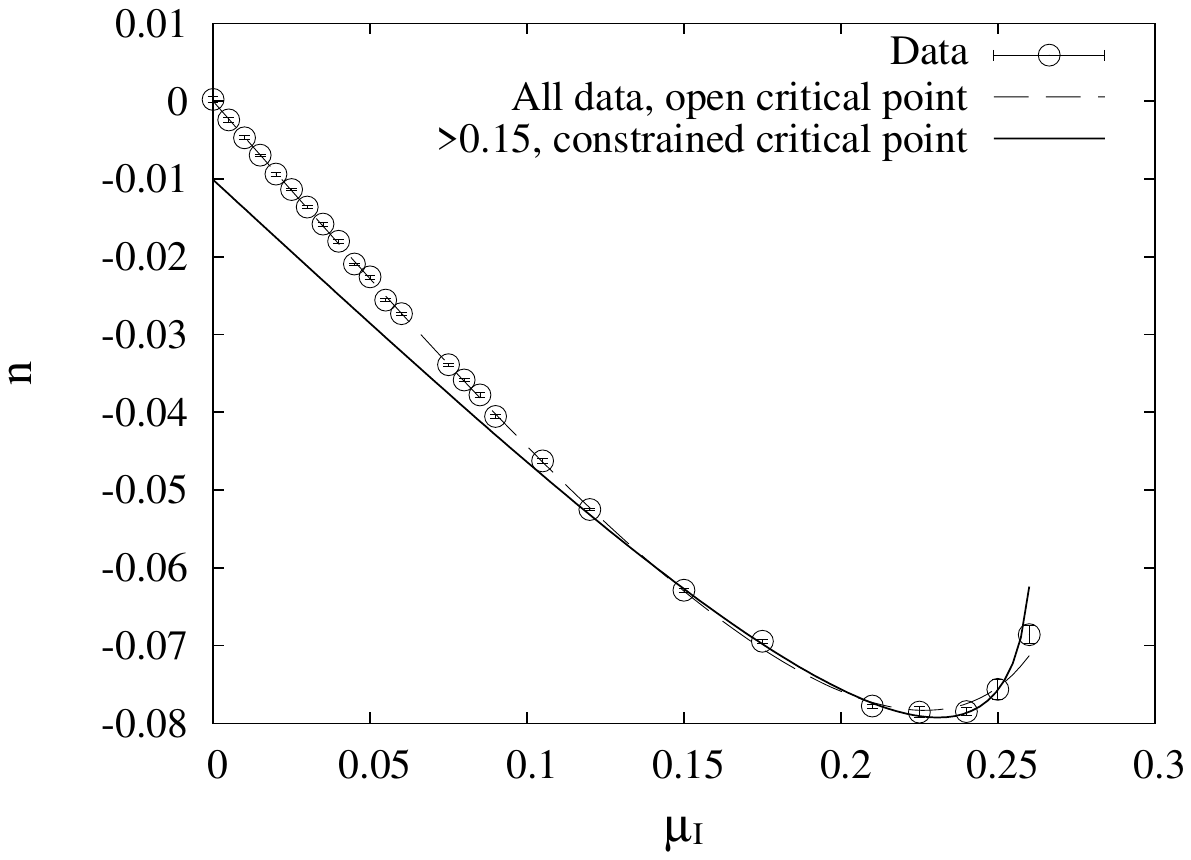}
\end{minipage}
\hskip -2.7 truecm
\begin{minipage}{11 truecm}
\vskip -0.5 truecm
\includegraphics[width=1.1\textwidth]{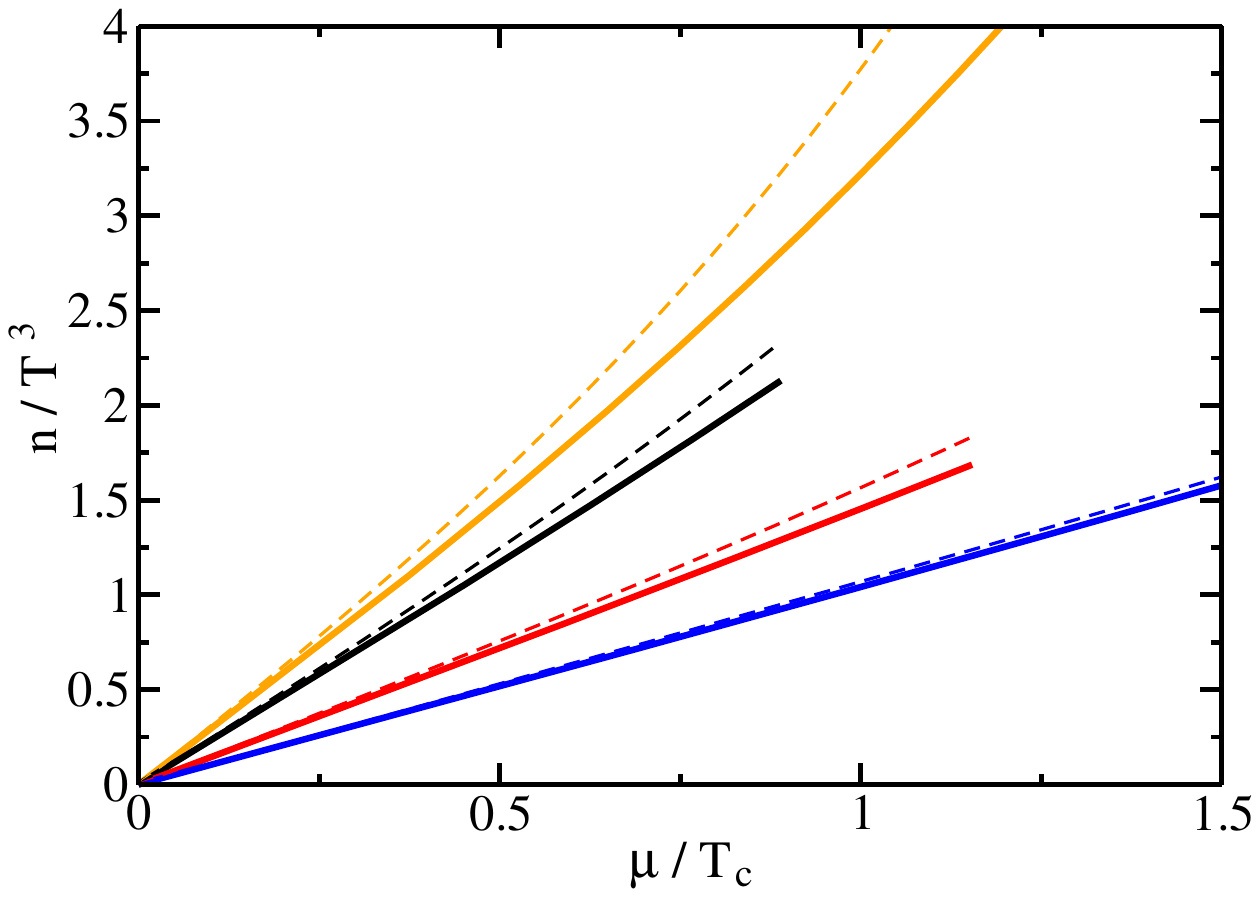}
\end{minipage}
\end{minipage}
\vskip -1.0 truecm
\caption{The particle density as a function of imaginary chemical potential,
at $T=1.1 T_c$ is well fitted by a conventional critical behaviour (left,
from ref. \cite{D'Elia:2007ke} ), 
as well as by a quasiparticle model with 
$\mu$ dependent coefficients\cite{Bluhm:2007cp}.
Once continued to real chemical potential, this 
gives a modified form of the Stefan-Boltzmann law 
(right, upper curve,  from \cite{Bluhm:2007cp}) , 
which by increasing T 
gradually approaches the free field behaviour, corresponding to the
lower curve in the plot}
\end{figure}
At nonzero chemical potential, it turns 
 out that it is useful to consider the phase diagram of QCD
in the $T,\mu^2$ plane. It has been found \cite{D'Elia:2007ke}
that the strongly interactive quark gluon plasma can be described 
by  $p(T, \mu) = b(T) |t + a(T) (\mu^2 + \mu_c^2)|^ {(2-\alpha)}$ , 
implying $ n(T, \mu) =  A(T) \mu ({\mu^c}^2 + \mu^2)^{(2-\alpha)}$,
where $\mu_c$ is the critical point at imaginary chemical potential, see
Fig. 5.
Note that the simple polynomial behaviour of the free field limit
would be recovered when $\alpha=1$, and that $\alpha > 1.$,
resulting from the numerical analysis, implies a slower increase
of the particle number in the critical region with respect to
the free case. It would be interesting to repeat this investigation
using the generalised method of ref.\cite {Azcoiti:2005tv}.
 Data at imaginary chemical potential are amenable
to an easy comparison with analytic studies, and indeed the results
in the strongly interactive region compares favourably with a quasiparticle
study\cite{Bluhm:2007cp} , once  an explicit dependence of the  
self-energy parts on $\mu_i = \mu_{u,d}$ and $T$, as well as an implicit 
dependence via the effective coupling $G^2(T, \mu_u, \mu_d)$ has
been taken into account:
\begin{equation}
\omega^2_i = k^2 + m_i^2 + \Pi_i, \quad 
\Pi_i = \frac13 \left(T^2 + \frac{\mu_i^2}{\pi^2} \right) G^2 (T,\mu_u,\mu_d)\,.\end{equation}
At larger temperature the data should compare successfully with a resummed
perturbation theory \cite{Vuorinen:2003fs}, and it remains to be seen which is the lower limit of its applicability. 
\section{Towards Fair}
While the results of the two previous sections are robust, and we believe
they can be systematically improved much in the same way as standard lattice
QCD calculations, in this colder, denser region the sign problem becomes
more severe and the results should still be considered exploratory
\cite{Lombardo:2004uy}. 

\begin{figure}[t]
\vskip -5 truecm
\begin{minipage}{11truecm}
\vskip 2 truecm
\hskip 0.3 truecm
\includegraphics[width=0.8\textwidth]{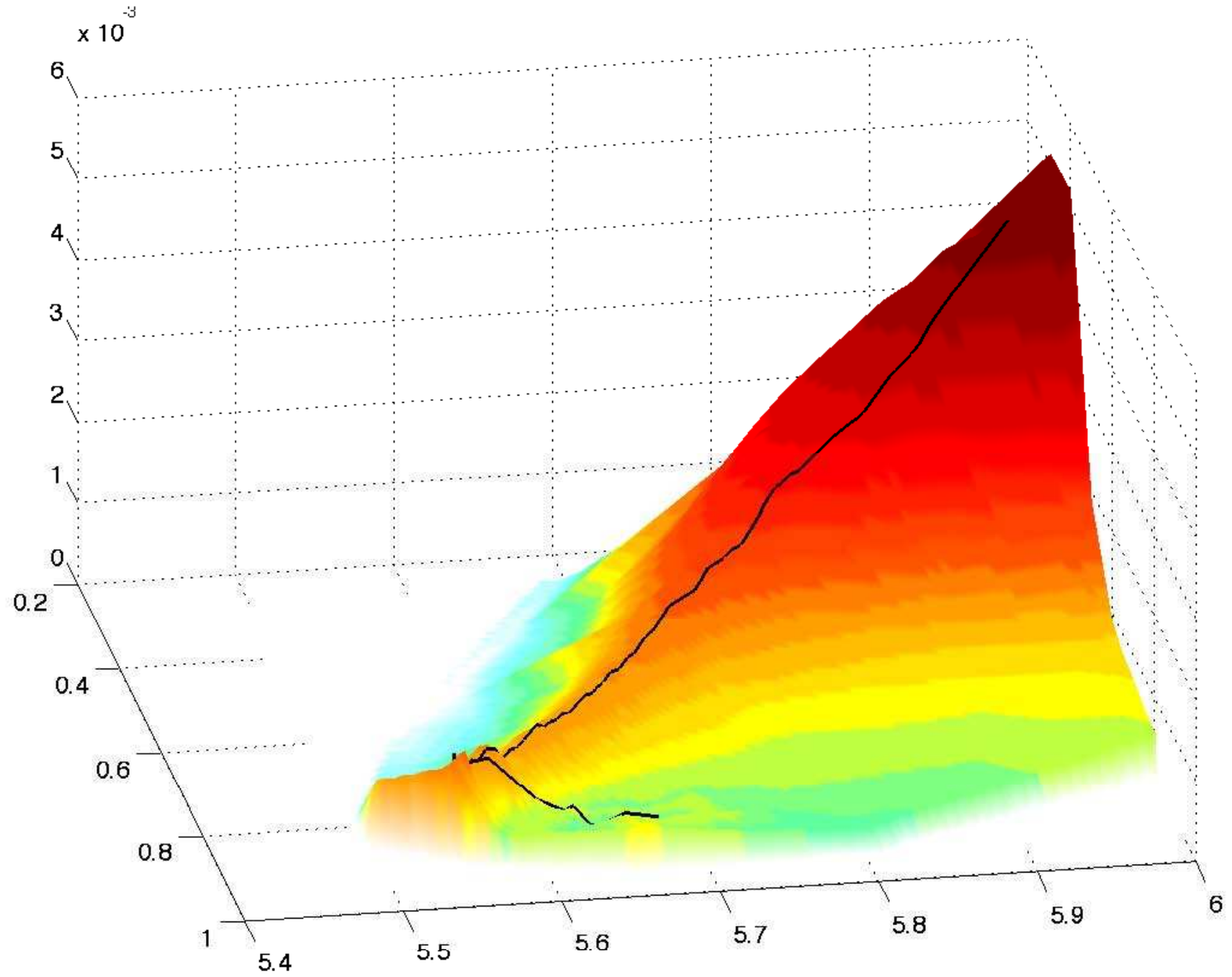}
\end{minipage}
\hskip -4. truecm
\begin{minipage}{11truecm}
\vskip -19 truecm
\includegraphics[width=2.1\textwidth]{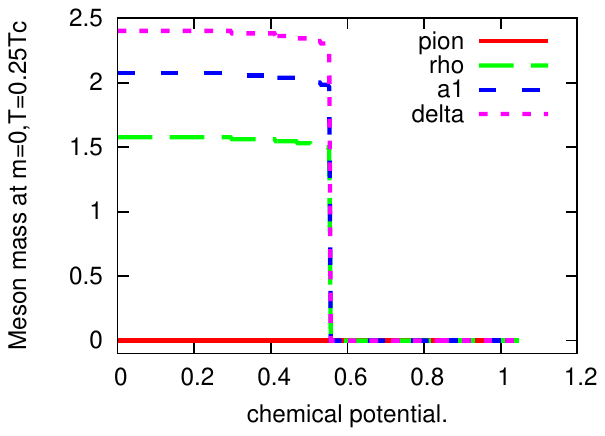}
\end{minipage}
\vskip -2 truecm
\caption{Results from QCD expansions:
The Polyakov loop as a function of T, $\mu$
from a numerical study at large mass(left \cite{DePietri:2007ak}) and 
the pattern of chiral symmetry as seen from the meson spectrum
at low temperature as a function of $\mu$ from strong coupling QCD
(right,\cite{Kawamoto:2007cc})}
\label{fig:ale}
\end{figure}

The mass spectrum has been computed via the QCD 
strong coupling expansion, observing
the expected signatures of chiral symmetry restoration, and a $\rho$ mass
decreasing with temperature\cite{Kawamoto:2007cc}.
 Calculations have
also been performed in the double limit: 
$ M \rightarrow \infty,\, \mu \rightarrow \infty,\,  
\zeta \equiv {\rm exp}\,(\mu - \ln M) $  fixed , corresponding to
an evolved `quenched
approximation' in the presence of charged matter. The order parameter has
been computed (note the ridge in the $T, \mu$ plane in Fig. 
\ref{fig:ale}, left diagram), allowing
the observation of the phase transition . There are also
emerging indication of a tricritical point, and studies of diquark are
in progress 
\cite{DePietri:2007ak}.Similar results come from a study using
overlap fermions at strong coupling \cite{Yu:2005eu}.

New results in the canonical formalism, in which the canonical
partition function was evaluated following  Hasenfratz-Toussaint\cite{HaTou}:
\begin{eqnarray}
\frac{Z_C(B,\beta)}{Z_{GC}(\beta_0=\beta,\mu=i \mu_{I_0})} 
&=&
 \langle \frac{1}{2\pi} \int_{-\pi}^{\pi} d \left( \frac{ \mu_I }{ T }\right) \; e^{-i 3 B \frac{ \mu_I }{ T }}
\frac{ \det(U;i \mu_I) }{ \det(U;i \mu_{I_0}) } \rangle_{\beta_0, i\mu_{I_0}}
\nonumber  \\& 
\equiv &\langle \frac{\hat Z_C(U;B)}{\det(U;i \mu_{I_0})}\rangle_{\beta_0, i\mu_{I_0}} 
\end{eqnarray}
allowed the identification of a first order line,  a coexistence
region, and the associated critical $\mu$ and critical densities,
for four-flavor QCD\cite{deForcrand:2006ec}.

A variant of the canonical approach, based on the evaluation of the
Grand Partition Function via a Taylor expansion
 produced results for $N_f=2$, indicating
a qualitative change at $T/T_c \simeq 0.8$, and a first order
line \cite{Ejiri:2007tk}. 

The density of states method - a reordering of the functional
integral based on the costrained partition function (which in
condensed matter parlance might be called a mesoscopic approach )-

\begin{equation}
\rho(x)
=
\int \mathcal{D}U\, g(U) \, \delta( \phi - x ).
\end{equation}
applied to the $N_f=4$ theory  allowed the identification of
two phase transition lines, and gave indication of a triple point
\cite{Fodor:2007vv} (Fig.7, left).

\begin{figure}[t] 
\vskip -6.0 truecm
\begin{minipage}{11 truecm}
\vskip -2.9 truecm
\includegraphics[width=1.1\textwidth]{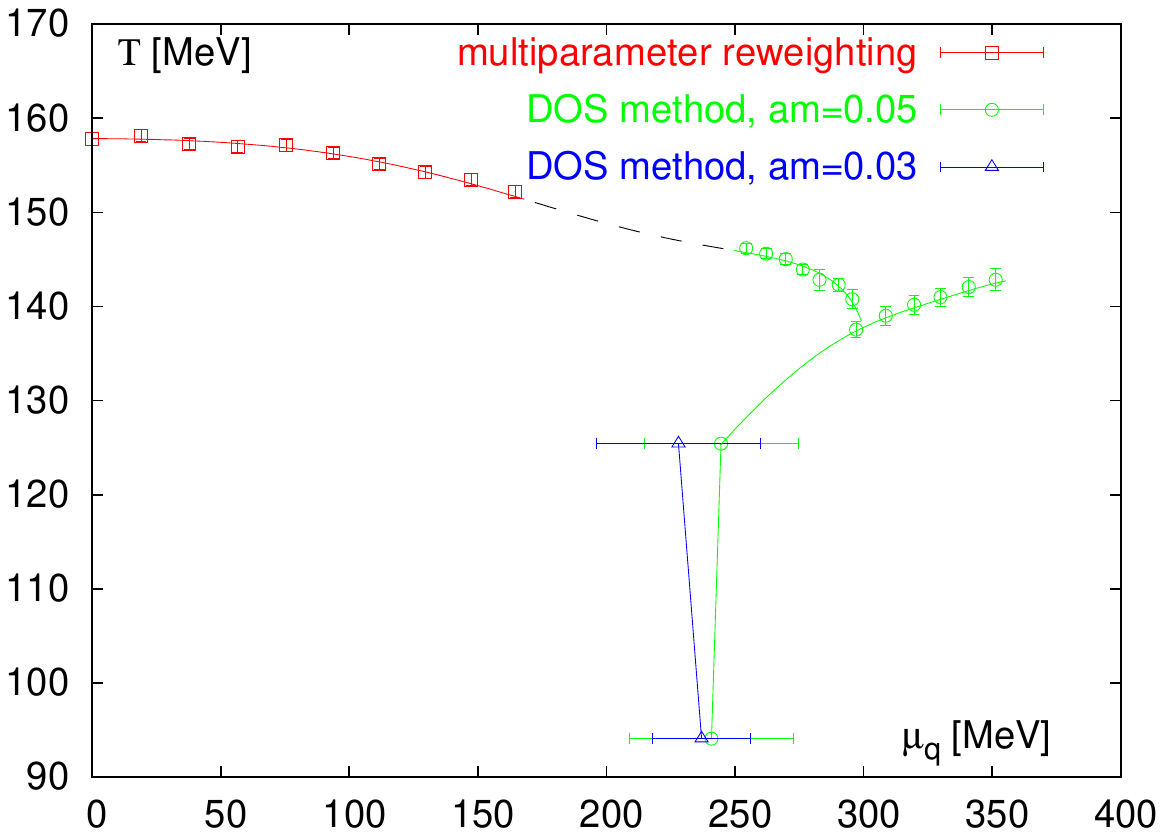}
\end{minipage}
\hskip -3 truecm
\begin{minipage}{11 truecm}
\vskip -0.1 truecm
\includegraphics[width=0.85\textwidth]{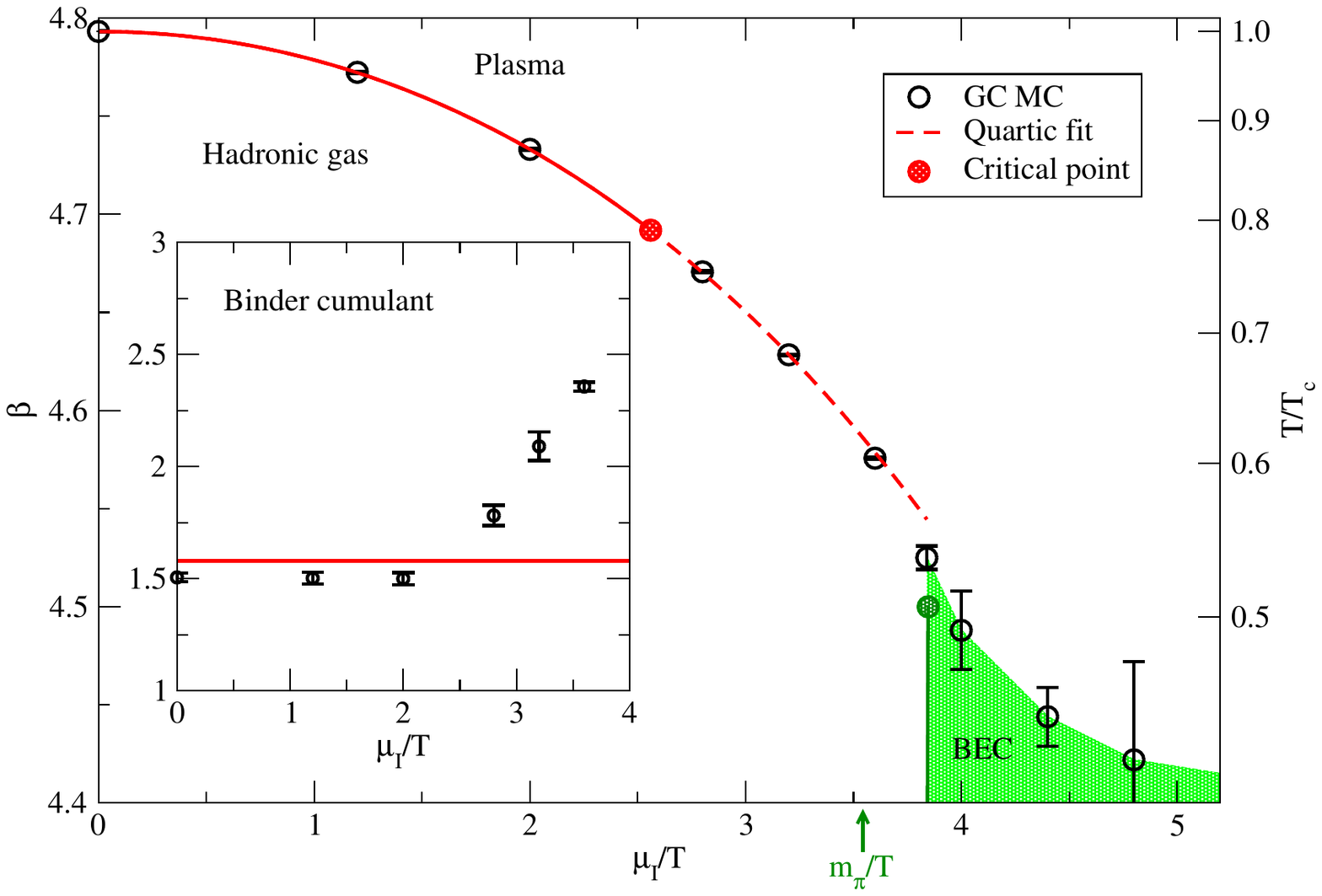}
\end{minipage}
\caption{The phase diagram of QCD in the T, $\mu_B$ plane,
from a (still exploratory)  Density of States calculation
(left diagram \cite{Fodor:2007vv}).
 The phase diagram of QCD in the $T, \mu_I$ plane (right diagram,
\cite{deForcrand:2007uz}).}
\end{figure}

\section{Cold and dense matter : QCD-like models}

Finally in the cold and dense phase, the sign problem is very
severe and at this moment I can only show result for 
QCD at non-zero density of isospin, and two-color QCD at nonzero
baryon density. These theories have extra symmetries, which 
protect the determinant from becoming complex, and allow
standard MonteCarlo simulation, see e.g ref. 
\cite{Hands:2007by}
for  reviews. 

 At zero temperature the onset
for thermodynamics equals, as usual, the mass of the lowest excitation
carrying the relevant charge, i.e. $\mu_c = m_\pi/2$ in the
case of isospin density, and $\mu_c = m_B/N_c = m_\pi/2$, for two-
color QCD, 
so the physics of a finite density of isospin, and of finite baryon density 
have many analogies.

The phase diagram of QCD in the $\mu_I, T$ plane has been recently
calculated, and the superfluid BEC phase has been clearly identified
\cite{deForcrand:2007uz} (Fig.7, right). 

Early studies, both analytic and numeric, have identified a superfluid
phase also in two-colour QCD, by inspecting the behaviour of fermionic
observables, and confirming the prediction of chiral perturbation
theory, which is applicable here
\cite{Hands:2007by}. 

There is an emerging consensus that the BEC phase of two-colour QCD is
still confining,and preliminary indications of a BEC/BCS crossover have been
reported \cite{Hands:2006ve}, with the
superfluid order parameter assuming the scaling 
consistent with Cooper pairing at a Fermi surface.
Results from ongoing work\cite{HKS:2008}  indicate that a BEC/BCS
transition persists at a finer lattice spacing. 
At the hadronic/BEC interface there are signal of criticality
in the gluonic sector, as well as non trivial
modifications of the glueball propagators in the superfluid phase
\cite{Lombardo:2007pw} (Fig.8).

\begin{figure}[t]
\vskip -6 truecm
\begin{minipage}{11 truecm}
\includegraphics[width=0.75\textwidth]{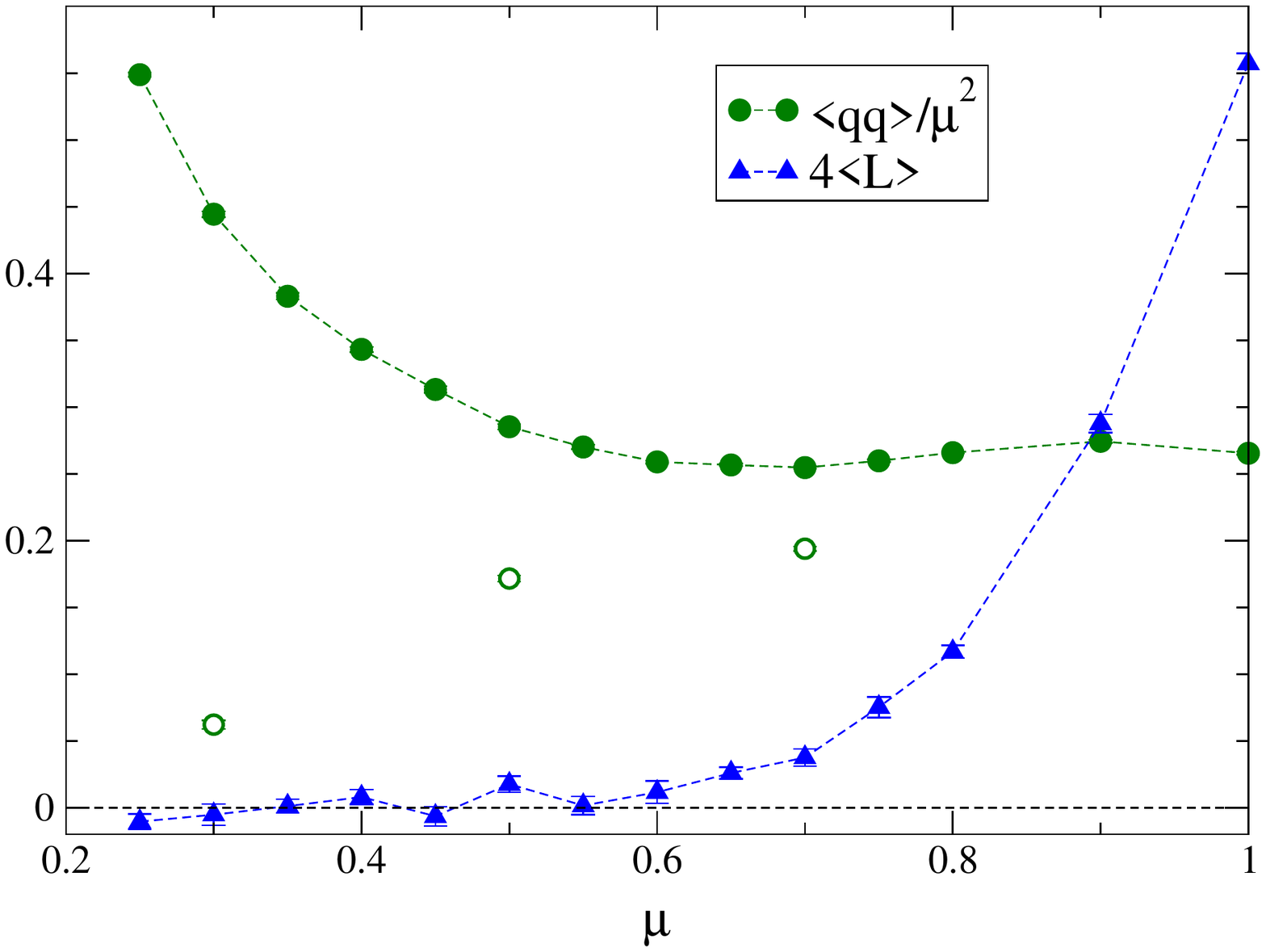}
\end{minipage}
\hskip -4.5 truecm
\begin{minipage}{11 truecm}
\vskip -5.7 truecm
\includegraphics[width=1.3\textwidth]{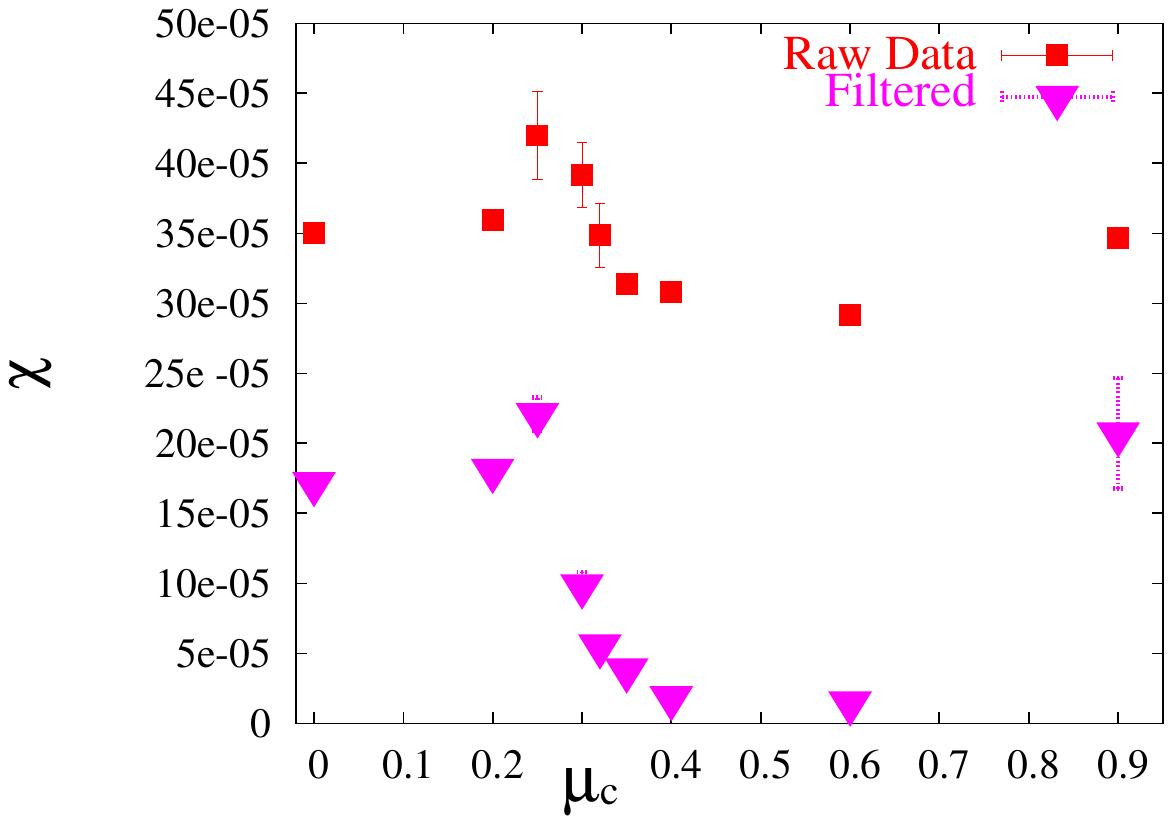}
\end{minipage}
\vskip -1 truecm
\caption{The superfluid phase of two colour QCD : 
  diquarks condense (open symbols are for <qq>
extrapolated to the chiral limit), however the Polyakov loop
remains close to zero,  hence the superfluid phase still confines
(left, from ref \cite{Hands:2006ve};
the amplitude of the glueball propagator peaks at the critical point,
and reaches a lower value above $\mu_c$, indicating non trivial
modification of the baryon-dense gluonic medium (right, from ref.
\cite{Lombardo:2007pw}).}
\label{fig:su2glue}
\end{figure}

One final comment concerns the large $N_c$ limit, where a scenario has
been proposed in which deconfinement and chiral transition are
separated\cite{McLerran:2007qj}. 
It is very tempting to speculate on possible analogies
of these observation with the behaviour of $N_c = 2$, where indeed
chiral transition (even if between two phases where chiral symmetry remains
broken) and deconfinement transition are different phenomena.

\ack
I wish to thank the Organisers of QM2008 
for a most pleasant and interesting Conference. I am very grateful
to  Gordon Baym,  Philippe de Forcrand, Massimo D'Elia, Francesco Di Renzo, 
Alessandra Feo, Zoltan Fodor,  
Christa Guse, Simon Hands, Tetsuo Hatsuda,  Frithjof  Karsch, Edwin Laermann, 
Larry McLerran, Owe Philipsen, Christian Schmidt,   Zsolt Szep, and 
Aleksi Vuorinen
for many helpful discussions and correspondence.

\end{document}